# Preference-based Graphic Models for Collaborative Filtering

**Rong Jin, Luo Si**
School of Computer Science
Carnegie Mellon University
Pittsburgh, PA 15232
{rong,lsi}@cs.cmu.edu

**ChengXiang Zhai**
Department of Computer Science
University of Illinois at Urbana-Champaign
Urbana, IL 61801
czhai@cs.uiuc.edu

**Abstract**

Collaborative filtering is a very useful general technique for exploiting the preference patterns of a group of users to predict the utility of items to a particular user. Previous research has studied several probabilistic graphic models for collaborative filtering with promising results. However, while these models have succeeded in capturing the similarity among users and items, none of them has considered the fact that users with similar interests in items can have very different rating patterns; some users tend to assign a higher rating to all items than other users. In this paper, we propose and study two new graphic models that address the distinction between user *preferences* and *ratings*. In one model, called the decoupled model, we introduce two different variables to *decouple* a user's preferences from his/her ratings. In the other, called the preference model, we model the *orderings* of items preferred by a user, rather than the user's numerical ratings of items. Empirical study over two datasets of movie ratings shows that, due to its appropriate modeling of the distinction between user preferences and ratings, the proposed decoupled model significantly outperforms all the five existing approaches that we compared with. The preference model, however, performs much worse than the decoupled model, suggesting that while explicit modeling of the underlying user preferences is very important for collaborative filtering, we can not afford ignoring the rating information completely.

## 1. INTRODUCTION

The rapid growth of the information on the Internet demands intelligent information agent that can sift through all the available information and find out the most valuable to us. These intelligent systems can be categorized into two classes: *Collaborative Filtering* (CF) and *Content-based recommending*. The difference between them is that collaborative filtering only utilizes the ratings of training users in order to predict ratings for test users while content-based recommendation systems rely on the contents of items for predictions. Therefore, collaborative filtering systems have advantages in an environment where the contents of items are not available due to either a privacy issue or the fact that contents are difficult for a computer to analyze. In this paper, we will only focus on the collaborative filtering problems.

Most collaborative filtering methods fall into two categories: Memory-based algorithms and Model-based algorithms [Breese et al. 1998]. Memory-based algorithms store rating examples of users in a training database, and in the predicting phase, they would predict a test user's ratings based on the corresponding ratings of the users in the training database that are similar to the test user. In contrast, model-based algorithms build models that can explain the training examples well and predict the ratings of test users using the estimated models. Both types of approaches have been shown to be effective for collaborative filtering.

In general, all collaborative filtering approaches assume that users with similar "tastes" would rate items similarly, and the idea of clustering is exploited in all approaches either explicitly or implicitly. Compared with memory-based approaches, model-based approaches provide a more principled way of performing clustering, and is also often much more efficient in terms of the computation cost at the prediction time. The basic idea of a model-based approach is to cluster items and/or training users into classes explicitly and predict ratings of a test user by using the ratings of classes that fit the best with the test user and/or items to be rated. Several different probabilistic models have been proposed and studied in the previous work (e.g., [Breese et al. 1998; Hofmann & Puzicha 1998; Pennock et al. 2000; Popescul et al. 2001; Ross & Zemel 2002] ). These models have succeeded in capturing user/item similarities through probabilistic clustering in one way or the other, and have all been shown to be quite promising. However, one common deficiency in all these previous models is that they are all based on the assumption that users with similar interests would rate items similarly, which is not true in reality. Indeed, the rating pattern of a user is determined not only by his/her interests but also by the rating strategy/habit. For example, some users are more "tolerate" than others, and therefore their ratings of items tend to be higher than others even though they share very similar tastes of items. This problem has already discussed in an early study of collaborative filtering by Resnick and others, and is often addressed through heuristic normalization in a memory-based approach [Resnick et al., 1994; Breese et al. 1998], but it has not been addressed in a model-based approach.

In this paper, we propose and study two new graphic models that address the distinction between user preferences and ratings. In one model, called the *decoupled model*, we introduce two different variables to *decouple* a user's preferences and ratings. Specifically, we use two hidden variables to account for the preferences (i.e., interests) and the rating patterns of a user, respectively. In the other model, called the *preference model*, we model the orderings of items preferred by a user, rather than the user's numerical ratings of items. The idea is to focus on the "essential" information conveyed by ratings, which is the implied relative preference orderings among items, so that the algorithm would not need to model the absolute rating values, which may be affected by a user's rating habit. For example, if a user gives items 'a', 'b' and 'c' a rating of 2, 3, 4 respectively, our preference model would only take it as meaning item 'c' is preferred to item 'b', which is preferred to item 'a'. This means that a rating of 3, 4, 5 for 'a', 'b', and 'c' would have precisely the same effect.

We evaluated these two models over two datasets of movie ratings. The results show that the decoupled model is quite successful in capturing the distinction between user preferences and ratings, and outperforms five existing approaches substantially and consistently. However, the preference model is not very successful. These results suggest that explicit modeling of the underlying user preferences is very important for collaborative filtering, but we can not afford ignoring the rating information completely.

The rest of paper is arranged as follows: Section 2 discusses previous work on model-based collaborative filtering. We present the two proposed graphic models in Section 3, and discuss the experiment results in Section 4. Conclusions and future work are discussed in Section 5.

## 2. MODEL-BASED CF APPROACHES

In this section, we briefly review existing probabilistic models for collaborative filtering and set up some technical background for our own models. First, let us introduce the annotations to be used in the rest of this paper. We let $X = \{x_1, x_2, ......, x_M\}$ be a set of items, $Y = \{y_1, y_2, ......, y_N\}$ be a set of users, and $\{1,...,R\}$ be a range of ratings. Let $\{(x_{(1)}, y_{(1)}, r_{(1)}), ......, (x_{(L)}, y_{(L)}, r_{(L)})\}$ be all the ratings in the training database and each tuple $(x_{(i)}, y_{(i)}, r_{(i)})$ means that user $x_{(i)}$ gives item $y_{(i)}$ a rating of $r_{(i)}$. For each user $y$, let $X(y)$ be the set of rated items, $R_y(x)$ be the rating of item $x$, and $\bar{R}_y$ be the average rating. We now discuss the three major probabilistic models for collaborative filtering: the Bayesian Clustering algorithm (BC) [Breese et al. 1998], the aspect model (AM) [Hofmann & Puzicha, 1999], and the Personality Diagnosis model (PD) [Pennock et al., 2000].

### 2.1 BAYESIAN CLUSTERING (BC)

The basic idea of BC is to assume that the same type of users would rate items similarly, and thus users can be grouped together into a set of user classes according to their ratings of items. Formally, given a user class 'C', the preferences regarding the various items expressed as ratings are independent, and the joint probability of user class 'C' and ratings of items can be written as the standard naive Bayes formulation:

$$P(C, r_1, r_2, ..., r_M) = P(C) \prod_{i=1}^{M} P(r_i | C) \quad (1)$$

Then, the joint probability for the rating patterns of user $y$, i.e. $\{R_y(x_1), R_y(x_2), ..., R_y(x_M)\}$, can be expanded as:

$$P(R_y(x_1), R_y(x_2), ..., R_y(x_M)) = \sum_C P(C) \prod_{i \in X(y)} P(R_y(x_i) | C) \quad (2)$$

The Expectation-Maximization (EM) algorithm can be used to cluster users. More details can be found in [Breese et al. 1998].

### 2.2 ASPECT MODEL (AM)

The aspect model is a probabilistic latent space model, which models individual preferences as a convex combination of preference factors [Hofmann & Puzicha 1999]. The latent class variable $z \in Z = \{z_1, z_2, ......, z_K\}$ is associated with each observation pair of a user and an item. The aspect model assumes that users and items are independent from each other given the latent class variable. Thus, the probability for each observation pair (x,y) is calculated as follows:

$$P(x, y) = \sum_{z \in Z} P(z) P(x | z) P(y | z) \quad (3)$$

where $P(z)$ is class prior probability, $P(x|z)$ and $P(y|z)$ are class-dependent distributions for items and users, respectively. Intuitively, this model means that the preference pattern of a user is modeled by a combination of typical preference patterns, which are represented in the distributions of $P(z)$, $P(x|z)$ and $P(y|z)$.

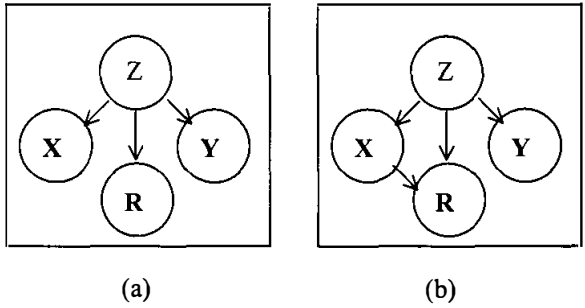

Figure 1: Graphical models for the two extensions of aspect model in order to capture rating values.

There are two ways to incorporate the rating information 'r' into the basic aspect model, which are expressed in Equation (4) and (5), respectively.

$$P(x_{(l)}, y_{(l)}, r_{(l)}) = \sum_{z \in Z} P(z) P(x_{(l)}|z) P(y_{(l)}|z) P(r_{(l)}|z) \quad (4)$$

$$P(x_{(l)}, y_{(l)}, r_{(l)}) = \sum_{z \in Z} P(z) P(x_{(l)}|z) P(y_{(l)}|z) P(r_{(l)}|z, x_{(l)}) \quad (5)$$

The corresponding graphical models are shown in Figure 1. The second model in Equation (5) has to estimate the conditional probability $P(r_{(l)}|z, x_{(l)})$, which has a large parameter space and may not be estimated reliably. Therefore, in our experiments, we will only compare the aspect model in Equation (4). (Equation (4) also actually performs better than (5).)

Unlike the Bayesian Clustering algorithm, where only the rating information is modeled, the aspect model is able to model the users and the items with conditional probability $P(y|z)$ and $P(x|z)$. Our decoupled model extends these aspect models by introducing additional hidden variables to model the preferences and ratings separately.

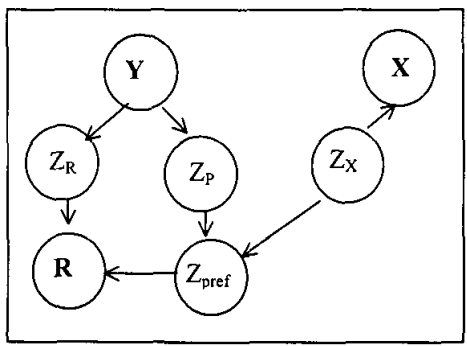

Figure 2: Graphical model representation for the decoupled models of preference and rating patterns

### 2.3 PERSONALITY DIAGNOSIS MODEL (PD)

In the personality diagnosis model, the observed rating for a test user $y'$ on an item $x$ is assumed to be drawn from an independent normal distribution with the mean as the true rating as $R_{y'}^{True}(x)$:

$$P(R_{y'}(x) | R_{y'}^{True}(x)) \propto e^{-(R_{y'}(x) - R_{y'}^{True}(x))^2 / 2\sigma^2} \quad (6)$$

where the standard deviation $\sigma$ is set to constant 1 in our experiments. Then, the probability of generating the observed rating values of the test user by any user $y$ in the training database can be written as:

$$P(R_{y'} | R_y) \propto \prod_{x \in X(y')} e^{-(R_y(x) - R_{y'}(x))^2 / 2\sigma^2} \quad (7)$$

The likelihood for the test user $y'$ to rate an unseen item $x$ as category $r$ can be computed as:

$$P(R_{y'}(x) = r) \propto \sum_y P(R_{y'} | R_y) e^{-(R_y(x) - r)^2 / 2\sigma^2} \quad (8)$$

The final predicted rating for item 'x' by the test user will be the rating category 'r' with the highest likelihood $P(R_{y'}(x) = r)$. Different from previous two approaches where users are clustered into user classes, this approach treats each user as a different model. Therefore, this approach is able to maintain the diversity of model ensemble. However, it may suffer significantly from data sparseness because most users only rate a small portion of items and a model based on a small number of ratings may be unreliable. Nevertheless, empirical studies have shown that this Personal Diagnosis method is able to outperform many other approaches for collaborative filtering [Pennock et al., 2000].

## 3. PREFERENCE-BASED MODELS

In this section, we discuss two new graphic models for collaborative filtering, both trying to make a distinction between the underlying preferences of a user on items and the surface item ratings given by a user. The first approach models preferences and ratings of users separately; while the second models the underlying relative rating patterns instead of the surface rating values.

### 3.1 DECOUPLED MODELS FOR RATING AND PREFERENCE PATTERNS (DM)

To account for the fact that users with similar interests may have very different rating patterns, we extend the aspect models by introducing two hidden variables $Z_P$, $Z_R$, with $Z_P$ for the preference patterns of users and $Z_R$ for the rating patterns of users. The whole model is shown in Figure 2. $Z_X$ is a class of items. Users are clustered from two different perspectives, i.e., $Z_P$ represents a grouping of users based on their preference patterns, whereas $Z_R$ groups users based on their rating patterns or habits. The domain sizes of $Z_X$, $Z_R$, and $Z_P$ are 5, 10, and 3 in our experiments. $Z_{pref}$ indicates whether or not the class of items $Z_X$ is preferred by the class of users $Z_P$ who presumably share similar preferences of items. The conditional probabilities $P(z_P | y_{(i)})$ and $P(z_R | y_{(i)})$ are the likelihood for user $y_{(i)}$ to be in the class of certain preference patterns $z_P$ and in the class of certain rating patterns $z_R$, respectively. Probabilities $P(z_X)$ and $P(x_{(l)} | z_X)$ are the priors of class of items $z_X$ and the likelihood of item $x_{(l)}$ to be generated from class $z_X$, respectively. Finally, $P(z_{pref} | z_P, z_X)$ is the probability for preference class $z_P$ to prefer item class $z_X$, and $P(r_{(l)} | z_R, z_{pref})$ is the likelihood for rating classes $z_R$ to rate items as $r_{(l)}$ given the preference condition $z_{pref}$.

We treat any tuple $(x_{(i)}, y_{(i)}, r_{(i)})$ as an observation of $(x_{(i)}, r)$ conditioned on $y_{(i)}$, and so its probability is

$$P(x_{(l)}, r_{(l)} | y_{(l)})
= \sum_{z_P, z_R, z_X} \left\{ \begin{array}{l} P(z_P | y_{(l)}) P(z_R | y_{(l)}) P(z_X) P(x_{(l)} | z_X) \times \\ \sum_{z_{pref}=0}^{1} P(z_{pref} | z_P, z_X) P(r_{(l)} | z_R, z_{pref}) \end{array} \right\} \quad (9)$$

We will call this model '**DM**'.

Comparing the 'DM' model in Figure 2 with the aspect model in Figure 1, we see that they differ in two aspects:

1) In 'DM', users and items are modeled separately, with hidden variable $Z_X$ for the clusters of items and hidden variables $Z_P$ and $Z_R$ for explaining the preference patterns and rating patterns of users, whereas in Figure 1, there is only one hidden variable $Z$ for describing classes of users and items.

2) The rating value is determined jointly by the hidden variables $Z_R$ and $Z_{pref}$. Therefore, even if a user likes a certain type of items, the rating value can still be low if he has a very 'tough' rating criterion. Thus, with the introduction of hidden variable $Z_R$, we are able to account for the variance in rating patterns among the users with similar interests.

Furthermore, several design issues need to be discussed:

1) In 'DM', we assume conditional independence between hidden variables $Z_P$ and $Z_R$ given the user $y$, which makes it possible to simplify the conditional probability $P(z_P, z_R|y)$ as a product of $P(z_P|y)$ and $P(z_R|y)$. This helps decrease the number of parameters significantly and avoid the problem of sparse data because the product space of $Z_P$ and $Z_R$ can be quite large. Furthermore, this choice makes the inference process computationally fast.

2) In Equation (9), we only consider two cases for the hidden variable $Z_{pref}$, namely $Z_{pref}=1$ for the user to prefer an item and $Z_{pref}=0$ for not preferring. In general, we can increase the preference levels of preferring items as we want. For example, we can have three different preference levels, with zero for no preference, one for slight preference and two for strong preference. In our experiments, we set the number of preference levels to the number of different rating categories.

### 3.1.1 The Training Procedure

With hidden variables in the model, the Expectation and Maximization (EM) algorithm is a natural choice [Dempster & Rubin 1977]. The EM algorithm alternates between the expectation steps and maximization steps. In the expectation step, the joint posterior probabilities of the latent variables $\{Z_P, Z_R, Z_X, Z_{pref}\}$ are computed as

$$P(z_P, z_R, z_X, z_{pref} | x_{(l)}, y_{(l)}, r_{(l)})
= \frac{\left\{ \begin{array}{l} P(z_X) P(x_{(l)} | z_X) P(z_P | y_{(l)}) P(z_R | y_{(l)}) \\ \times P(z_{pref} | z_P, z_X) P(r_{(l)} | z_R, z_{pref}) \end{array} \right\}}{\sum_{z_P, z_R, z_X, z_{pref}} \left\{ \begin{array}{l} P(z_X) P(x_{(l)} | z_X) P(z_P | y_{(l)}) P(z_R | y_{(l)}) \times \\ P(z_{pref} | z_P, z_X) P(r_{(l)} | z_R, z_{pref}) \end{array} \right\}} \quad (10)$$

In the maximization step, the model parameters can be updated as follows:

$$P(z_X) = \frac{\sum_l \sum_{z_P, z_R, z_{pref}} P(z_X, z_P, z_R, z_{pref} | x_{(l)}, y_{(l)}, r_{(l)})}{L} \quad (11)$$

$$P(x|z_X) = \frac{\sum_{l:x_{(l)}=x} \sum_{z_P, z_R, z_{pref}} P(z_X, z_P, z_R, z_{pref} | x_{(l)}, y_{(l)}, r_{(l)})}{L \times P(z_X)} \quad (12)$$

$$P(z_P|y) = \frac{\sum_{l:y_{(l)}=y} \sum_{z_R, z_{pref}, z_X} P(z_X, z_P, z_R, z_{pref} | x_{(l)}, y_{(l)}, r_{(l)})}{\sum_{l:y_{(l)}=y} \sum_{z_P, z_R, z_{pref}, z_X} P(z_X, z_P, z_R, z_{pref} | x_{(l)}, y_{(l)}, r_{(l)})} \quad (13)$$

$$P(z_R|y) = \frac{\sum_{l:y_{(l)}=y} \sum_{z_P, z_{pref}, z_X} P(z_X, z_P, z_R, z_{pref} | x_{(l)}, y_{(l)}, r_{(l)})}{\sum_{l:y_{(l)}=y} \sum_{z_P, z_R, z_{pref}, z_X} P(z_X, z_P, z_R, z_{pref} | x_{(l)}, y_{(l)}, r_{(l)})} \quad (14)$$

$$P(z_{pref}|z_P, z_X) = \frac{\sum_l \sum_{z_R} P(z_X, z_P, z_R, z_{pref} | x_{(l)}, y_{(l)}, r_{(l)})}{\sum_l \sum_{z_R, z_{pref}} P(z_X, z_P, z_R, z_{pref} | x_{(l)}, y_{(l)}, r_{(l)})} \quad (15)$$

$$P(r|z_R, z_{pref}) = \frac{\sum_{l:r(l)=r} \sum_{z_P, z_X} P(z_X, z_P, z_R, z_{pref} | x_{(l)}, y_{(l)}, r_{(l)})}{\sum_l \sum_{z_P, z_X} P(z_X, z_P, z_R, z_{pref} | x_{(l)}, y_{(l)}, r_{(l)})} \quad (16)$$

In order to avoid bad local minimums, we use annealing EM [Ueda et al., 1998].

### 3.1.2 The Prediction Procedure

To predict the ratings of items by a test user $y^t$, we need to compute the probability distribution over preference classes $P(Z_P | y^t)$ and rating classes $P(Z_R | y^t)$. Let $X(y^t) = \{(x^t_{(1)}, y^t, r^t_{(1)}), ..., (x^t_{(N\_G)}, y^t, r^t_{(N\_G)})\}$ be the set of items that have been previously rated by $y^t$. Then, the optimal distributions $P(Z_P | y^t)$ and $P(Z_R | y^t)$ can be found by maximizing the log-likelihood of the items rated by $y^t$, i.e.

$$L(y') = \sum_{i=1}^{N\_G} \log P(r'_{(i)}, x'_{(i)} | y') \qquad (17)$$

where, $P(r'_{(i)}, x'_{(i)} | y')$ is computed using Equation (9) with all other parameters computed from training except for $P(Z_P | y')$ and $P(Z_R | y')$, which we will estimate. In order to make the estimated distributions less skewed, Laplacian smoothing is applied within the EM algorithm, which is the same as adding a Dirichlet prior of uniform mean on distributions $P(Z_P | y')$ and $P(Z_R | y')$.

## 3.2 MODELING PREFERRED ORDERING OF ITEMS (MP)

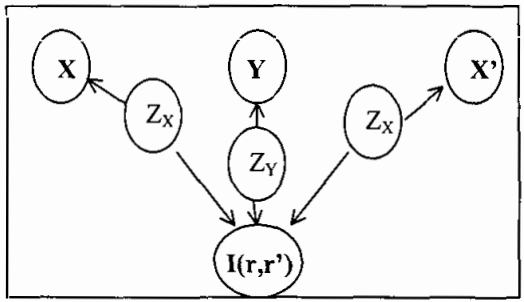

Figure 3: Graphic model representation for only modeling the relative orderings

In this subsection, we present another approach to addressing the variances in the rating patterns of the users with similar interests, which is to model the relative ordering between items instead of the absolute rating values. For example, if a user rates item 'a' as 2 and item 'b' as 3, we would only take it as meaning that 'b' is preferred to 'a'. That is, we assume that the relative orderings between items 'a' and 'b' are more consistent than the absolute values of ratings within the class of users with similar interests. Formally, let $I(r,r')$ be an indicator function for ratings $r$ and $r'$, defined as follows:

$$I(r,r') = \begin{cases} 0 & r = r' \\ 1 & r > r' \\ 2 & r < r' \end{cases} \qquad (18)$$

Joint probability $P(y, x, x', I(R_y(x), R_y(x')))$, i.e. the probability for user $y$ to rate item $x$ relative item $x'$ by the order of $I(R_y(x), R_y(x'))$, can be written as

$$P(y, x, x', I(R_y(x), R_y(x')))$$
$$= \sum_{z_Y, z_X, z'_X} \begin{Bmatrix} P(y|z_Y)P(z_Y)P(x|z_X)P(z_X)P(x'|z'_X) \\ \times P(z'_X)P(I(R_y(x), R_y(x'))|z_X, z'_X, z_Y) \end{Bmatrix} \qquad (19)$$

where, hidden variables $Z_Y$, $Z_X$ are the class of users and items, respectively. The corresponding graphic model is illustrated in Figure 3. We call this proposed model 'MP'. $P(I(R_y(x), R_y(x'))|z_X, z'_X, z_Y)$ can be simplified as:

$$P(I|z_X, z'_X, z_Y) = \begin{cases} v(z_X, z_Y)v(z'_X, z_Y) + \\ (1 - v(z_X, z_Y))(1 - v(z'_X, z_Y)) & r = r' \\ v(z_X, z_Y)(1 - v(z'_X, z_Y)) & r > r' \\ (1 - v(z_X, z_Y))v(z'_X, z_Y) & r < r' \end{cases} \qquad (20)$$

Where, $v(z_X, z_Y)$ is the likelihood for user class $z_Y$ to prefer the item class $z_X$. The model can be trained by maximizing the probability of the relative orderings between all pairs of rated items. Similar to the algorithm presented in the last section, an EM algorithm can also be used to train the 'MP' model.

The prediction phase of this model is a bit difficult due to a lack of modeling the absolute rating information explicitly. Thus, instead of predicting the ratings directly, we first find the most appropriate ranking position for a testing item with respect to the items with known ratings and then infer the most likely rating from the relative ranking and the known rating values. For example, suppose the ratings of items 'a', 'b' and 'c' are given as 2, 4, and 5, and according to the trained model, we find that a testing item 'd' is preferred to item 'a' and less favored than item 'b' and 'c'. Then, the most appropriate rating for the testing item 'd' would be 3. By following this idea, the most appropriate rating for a testing item $x'$ is found by maximizing the likelihood for the rating of $x'$ to be consistent with ratings of given items of a testing user, i.e.

$$r^* = \arg\max_{r'} \prod_{x \in X(y')} P(x, x', y', I(r', R_{y'}(x))) \qquad (21)$$

Where, $P(x, x', y', I(r', R_{y'}(x)))$ can be computed using Equation (19). The distribution $P(y'|z_Y)$ is pre-required for the computation of the most likely rating for a testing item. Similar to the procedure discussed in the previous subsection, we can compute $P(y'|z_Y)$ by maximizing the probability for all the relative orderings of the items rated by the testing user $y'$. One disadvantage of this prediction procedure is the existence of ratings with tied probability. For example, if two items are rated by the testing user as 3 and 4, and the testing item is found to be less favored than both these items, then, both rate category 1 and 2 would be consistent with the given ratings, and it would be impossible to find out which rating is more likely. In the experiment, we break such a tie by choosing the rating that is closer to the mean of given ratings.

Unlike the 'DM' model, where the preference information of users is explicitly separated from the rating information through two sets of hidden variables, the 'MP' model obtains the preference information of users by considering the relative rating between items, which makes this model relatively simple in the training phase but considerably difficult in the prediction phase. More specifically, due to the fact that this approach does not model absolute values of rating, the prediction of rating values could be inaccurate compared to the previously proposed approach.

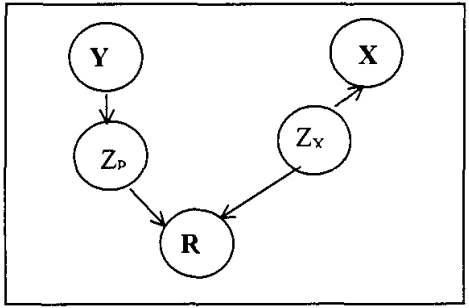

**Figure 4**: Graphical model representation for the baseline model for model 'DM'

## 4. EXPERIMENTS

In this section, we will present experiment results in order to address the following three issues:

1) Which of the two graphic models that we propose is better? Model 'MP' has the potential problem with predicting the rating values because it does not model absolute values of rating, whereas model 'DM' is considerably complicated and may have many local minimums over the surface of its log-likelihood function. We want to see which one performs better.

2) Would modeling the distinction between the preferences and ratings help improve the performance? In order to see the effectiveness of the 'DM' model, we will introduce a baseline model, which is almost identical to 'DM' except for the hidden nodes $Z_R$ and $Z_{pref}$. This baseline model is illustrated in Figure 4. It differs from the 'DM' model in that it infers the rating values directly from the class of preference $Z_p$ while 'DM' would infer the likelihood of preference $Z_{pref}$ first and then apply its rating class $Z_R$ to decide the rating value.

3) How effective are the proposed models compared with the previously proposed models? In this experiment, we will compare 'DM' and 'MP' with the three major previously studied graphic models and two memory-based approaches. In previous studies, when compared with the memory-based approaches, the model-based approaches tend to have mixed results [Breese et al. 1998]. It is thus interesting to see if our models, which decouple the preference patterns from rating patterns, can outperform memory-based approaches.

Two datasets of movie ratings are used in our experiments, i.e., 'MovieRating'[1] and 'EachMovie'[2]. A major challenge in collaborative filtering applications is for the system to operate effectively when it has not yet acquired a large amount of training data (i.e., the so-called "cold start" problem). To test our algorithms in such a challenging and realistic scenario, we decided to use only a subset of users from 'EachMovie', since we can reasonably expect any algorithm to perform better as we have more training data. Specifically, we extracted a subset of 2,000 users with more than 40 ratings from 'EachMovie'. The global statistics of these two datasets as used in our experiments are summarized in Table 1.

To compare algorithms more thoroughly, we experimented with several different configurations. For MovieRating, we take the first 100 or 200 users for training and all others for testing, whereas for EachMovie we use the first 200 or 400 users for training and the rest for testing. Furthermore, for each testing user, we varied the number of items with given ratings (5, 10, and 20). By varying the number of users for training, we can test the robustness of the learning procedure, and with different number of given items, we can test the robustness of the prediction procedure. In all experiments, the domain sizes of Zx, ZR, and ZP are set to 5, 10, and 3. We tried a few other values, and found that all turned out with similar performance.

The evaluation metric used in our experiments was the mean absolute error (MAE), which is the average absolute deviation of the predicted ratings to the actual ratings on items the test user has actually voted.

$$MAE = \frac{1}{L_{Test}} \sum_{l} | r_{(l)} - \hat{R}_{y_{(l)}}(x_{(l)}) | \qquad (16)$$

where $L_{Test}$ is the number of the test ratings.

**Table 1**: Characteristics of MovieRating and EachMovie.

|  | MovieRating | EachMovie |
|---|---|---|
| Number of Users | 500 | 2000 |
| Number of Items | 1000 | 1682 |
| Avg. # of rated Items/User | 87.7 | 129.6 |
| Number of Ratings | 5 | 6 |

**Table 2**: MAE of 'DM' and 'MP' on MovieRating. A smaller value means a better performance.

| Training Users Size | Algorithms | 5 Items Given | 10 Items Given | 20 Items Given |
|---|---|---|---|---|
| 100 | DM | 0.814 | 0.810 | 0.799 |
|  | MP | 0.911 | 0.905 | 0.880 |
| 200 | DM | 0.790 | 0.777 | 0.761 |
|  | MP | 0.877 | 0.861 | 0.837 |

### 4.1 EXPERIMENT 1: COMPARE 'DM' WITH 'MP'

---

[1] http://www.cs.usyd.edu.au/~irena/movie_data.zip

[2] http://research.compaq.com/SRC/eachmovie

The MAE results for both 'DM' and 'MP' over the tw testbeds with six different configurations are presented in Table 2 and 3. Clearly, the 'DM' model outperforms the 'MP' model in all cases. This is somehow expected given that 'MP' only models the relative orderings of items and therefore can have problems with predicting the absolute values of ratings. A more appropriate evaluation of models such as 'MP' should be based on preference relations rather than the ratings, which is clearly an important future work.

Table 3: MAE of 'DM' and 'MP' on EachMovie. A smaller value means a better performance.

| Training Users Size | Algorithms | 5 Items Given | 10 Items Given | 20 Items Given |
|---|---|---|---|---|
| 200 | DM | **1.07** | **1.04** | **1.03** |
|  | MR | 1.12 | 1.09 | 1.09 |
| 400 | DM | **1.05** | **1.03** | **1.02** |
|  | MP | 1.10 | 1.08 | 1.07 |

### 4.2 EXPERIMENT 2: COMPARE 'DM' WITH BASELINE PEERS

Table 4: MAE of 'DM' and its baseline peer on MovieRating. A smaller value means a better performance.

| Training Users Size | Algorithms | 5 Items Given | 10 Items Given | 20 Items Given |
|---|---|---|---|---|
| 100 | DM | **0.814** | **0.810** | **0.799** |
|  | Baseline | 0.823 | 0.822 | 0.817 |
| 200 | DM | **0.790** | **0.777** | **0.761** |
|  | Baseline | 0.804 | 0.801 | 0.799 |

Table 5: MAE of 'DM' and its baseline peer on EachMovie. A smaller value means a better performance.

| Training Users Size | Algorithms | 5 Items Given | 10 Items Given | 20 Items Given |
|---|---|---|---|---|
| 200 | DM | **1.07** | **1.04** | **1.03** |
|  | Baseline | 1.08 | 1.06 | 1.05 |
| 400 | DM | **1.05** | **1.03** | **1.02** |
|  | Baseline | 1.06 | 1.05 | 1.04 |

In this experiment, we compare the model 'DM' to its baseline peer, which is illustrated in Figure 4. These two models have exactly the same setup except that the model 'DM' introduces the extra hidden nodes $Z_R$ and $Z_{pref}$ in order to account for the variance in the rating behavior among the users with similar interests. The results are shown in Table 4 and 5. 'DM' outperforms the baseline model in all cases. Although the difference appears to be insignificant, it is interesting to note that when the number of rated items given increases, the gap between 'DM' and the baseline model also increases. This may suggest that when there are only a small number of items with given ratings, it is rather difficult to determine the type of rating patterns for the testing user. As the number of given items increases, this ambiguity will decrease quickly and therefore the advantage of the 'DM' model over the baseline model will be more clear. Indeed, it is a bit surprising that even with only five rated items and only a couple of hundreds of users the 'DM' model still slightly improves the performance as 'DM' has many more parameters to learn than the baseline model. We suspect that the skewed distribution of ratings among items, i.e., a few items account for a large number of ratings, may have helped.

### 4.3 EXPERIMENT 3: COMPARE 'DM', 'MP', WITH OTHER APPROACHES

Table 6: Comparison of 'DM', 'MP', and other existing methods in terms of MAE on MovieRating. A smaller value means a better performance.

| Training Users Size | Algorithms | 5 Items Given | 10 Items Given | 20 Items Given |
|---|---|---|---|---|
| 100 | PCC | 0.881 | 0.832 | 0.809 |
|  | VS | 0.859 | 0.834 | 0.823 |
|  | PD | 0.839 | 0.826 | 0.818 |
|  | AM | 0.882 | 0.856 | 0.836 |
|  | BC | 0.968 | 0.946 | 0.941 |
|  | DM | **0.814** | **0.810** | **0.799** |
|  | MP | **0.911** | **0.905** | **0.880** |
| 200 | PCC | 0.878 | 0.828 | 0.801 |
|  | VS | 0.862 | 0.950 | 0.854 |
|  | PD | 0.835 | 0.816 | 0.806 |
|  | AM | 0.891 | 0.850 | 0.818 |
|  | BC | 0.949 | 0.942 | 0.912 |
|  | DM | **0.790** | **0.777** | **0.761** |
|  | MP | **0.877** | **0.861** | **0.837** |

In this subsection, we compare both our models to other methods for collaborative filtering, including the Bayesian Clustering algorithm (BC), the Aspect Model (AM), the Personal Diagnosis (PD), the Vector Similarity method (VS) and the Pearson Correlation Coefficient method (PCC). The results are shown in Table 6 and 7.

The proposed model 'DM' is substantially better than all existing methods for collaborative filtering including both memory-based approaches and model-based approaches. These results are very promising, since they suggest that, compared with the memory-based approaches, graphic models are not only advantageous in principle, but also empirically superior due to their capabilities of capturing the distinction between the preference patterns and rating patterns in a principled way.

On the other hand, the MP model has mixed results; it is better than all existing methods on EachMovie, but mostly worse on MovieRating. This suggests that when the desired output is an absolute value, we can not afford to ignore the rating values completely, even though it is desirable to decouple the preference patterns from the rating patterns. In our future work, we will evaluate the MP model based on the preference relations, which would allow us to see more clearly its effectiveness.

Table 7: Comparison of 'DM', 'MP', and other existing methods in terms of MAE on EachMovie. A smaller value means a better performance.

| Training Users Size | Algorithms | 5 Items Given | 10 Items Given | 20 Items Given |
|---|---|---|---|---|
| 200 | PCC | 1.22 | 1.16 | 1.13 |
| | VS | 1.25 | 1.24 | 1.26 |
| | PD | 1.19 | 1.16 | 1.15 |
| | AM | 1.27 | 1.18 | 1.14 |
| | BC | 1.25 | 1.22 | 1.17 |
| | DM | **1.07** | **1.04** | **1.03** |
| | MR | **1.12** | **1.09** | **1.09** |
| 400 | PCC | 1.22 | 1.16 | 1.13 |
| | VS | 1.32 | 1.33 | 1.37 |
| | PD | 1.18 | 1.16 | 1.15 |
| | AM | 1.28 | 1.19 | 1.16 |
| | BC | 1.17 | 1.15 | 1.14 |
| | DM | **1.05** | **1.03** | **1.02** |
| | MP | **1.10** | **1.08** | **1.07** |

## 5. CONCLUSIONS AND FUTURE WORK

In this paper, we studied two different graphic models for collaborative filtering. Particularly, we focus ourselves on the problem that users with similar interests can have very different rating patterns, and proposed two different graphic models that can address this issue. The proposed 'DM' model avoids the variance in rating patterns by decoupling the rating patterns from the preference patterns, while the 'MP' model tries to achieve a similar effect by modeling the relative orderings of items instead of the absolute values of ratings.

Empirical results show that 'DM' is consistently better than 'MP' by the MAE measure, which is somehow expected due to the way the 'MP' model is designed. Furthermore, the experiments confirmed that the decoupling of rating patterns and preference patterns is important for collaborative filtering, and modeling such a decoupling in a graphic model leads to improvement in performance. Comparison with other methods for collaborative filtering indicates that the proposed method is superior, suggesting advantages of graphic models for collaborative filtering.

The idea of modeling preferences has also been explored in some other related work [Ha & Haddawy 1998; Freund et al. 1998;Cohen et al. 1999]. We plan to further explore this direction by considering all these different approaches and using a more appropriate evaluation criterion such as one based on inconsistent orderings. We also believe that the decoupling problem that we addressed may represent a more general need of modeling "noise" in similar problems such as gene microarray data analysis in bininformatics. We plan to explore a more general framework for all these similar problems.